\documentclass[letterpaper, 10 pt, conference]{ieeeconf}  
\IEEEoverridecommandlockouts
\usepackage{graphicx} 

\usepackage{amsmath}
\usepackage{amsfonts}
\usepackage{amssymb}
\usepackage{amsthm}
\usepackage{xcolor}
\usepackage{caption}
\usepackage{subcaption}
\usepackage{cite}
\theoremstyle{plain}
\newtheorem{thm}{Theorem}[section]
\newtheorem{lem}[thm]{Lemma}
\newtheorem{assump}[thm]{Assumption}

\title{\LARGE \bf
Comparing Run Time Assurance Approaches \\ for Safe Spacecraft Docking
}

\author{Kyle Dunlap,$^{1}$ Michael Hibbard,$^{2}$ Mark Mote,$^{3}$ and Kerianne Hobbs$^{4}$
\thanks{*Approved for Public Release, Case Numbers MSC/PA-2021-0345; AFRL-2021-2650. This work was supported by the Space Research Institute for Discovery and Exploration Fellowship, and the Air Force Research Laboratory Innovation Fund. The views expressed are those of the authors and do not reflect the official guidance or position of the United States Government, the Department of Defense or of the United States Air Force.}
\thanks{$^{1}$Kyle Dunlap is with the Department of Aerospace Engineering and Engineering Mechanics, University of Cincinnati, Cincinnati, OH, 45221,
        {\tt\small dunlapkp@mail.uc.edu}}%
\thanks{$^{2}$Michael Hibbard is with the Department of Aerospace Engineering and Engineering Mechanics, University of Texas at Austin,
        Austin, TX, 78712,
        {\tt\small mwhibbard@utexas.edu}}%
\thanks{$^{3}$Mark Mote is with the School of Aerospace Engineering, Georgia Institute of Technology,
        Atlanta, GA, 30332,
        {\tt\small marklmote@gmail.com}}%
\thanks{$^{4}$Kerianne Hobbs is on the Autonomy Capability Team (ACT3) at the Air Force Research Laboratory, Wright-Patterson AFB, OH, 45433,
        {\tt\small kerianne.hobbs@us.af.mil}}%
}

\begin{document}

\maketitle
\thispagestyle{empty}
\pagestyle{empty}

\begin{abstract}

Run Time Assurance (RTA) systems are online safety verification techniques that filter the output of a primary controller to assure safety. RTA approaches are used in safety-critical control to intervene when a performance-driven primary controller would cause the system to violate safety constraints. This paper presents four categories of RTA approaches based on their membership to explicit or implicit monitoring and switching or optimization interventions. To validate the feasibility of each approach and compare computation time, four RTAs are defined for a three-dimensional spacecraft docking example with safety constraints on velocity. 
\end{abstract}

\section{INTRODUCTION}

Designing control systems with high safety assurance can result in conservative designs that limit performance at the price of safety. One way to enforce safety of a system is the use of Run Time Assurance (RTA), which is an online safety assurance technique. Fundamentally, RTA approaches filter potentially unsafe inputs from a primary controller in a way that preserves safety of the system when necessary. The primary controller can range from a human operator to an autonomous control approach. RTA systems are designed to be independent from the structure of the primary controller, so that they can be applied to any system to assure safety. This independence allows a designer to decouple performance objectives and safety assurance.

This paper focuses on two different classes of RTA monitoring approaches, explicit or implicit, as well as two different classes of intervention approaches, switching-based or optimization-based. First, explicit approaches precisely define a safe set considering all constraints and may or may not use a backup controller. Implicit approaches use a predefined backup control law to compute trajectories and evaluate when to intervene \cite{gurriet2019scalable}. Explicit approaches are beneficial because they require less online computation, while implicit approaches are beneficial because they do not require a precise safe set to be defined, which can be overly conservative or impossible to concisely specify for complex systems.
In terms of intervention, switching-based approaches simply switch between a primary and backup control signal, and are sometimes called a simplex architecture \cite{rivera1996architectural}. Optimization-based intervention approaches use barrier constraints to minimize deviation from the primary control signal while assuring safety \cite{gurriet2020applied}. Switching approaches are simple, computationally efficient, and rely on backup control designs that meet safety and human-machine teaming constraints. On the other hand, optimization approaches are minimally invasive to the primary controller, where the intervention is smoother and more gradual than switching approaches.

In this paper, an RTA strategy is presented that is designed in the context of graceful degradation to have multiple phases of intervention. All systems begin with a \textit{Plan A} primary controller, which may or may not be verified. \textit{Plan B} RTA filters can use any combination of explicit, implicit, switching, and optimization approaches to actively assure safety. This phase of RTA is considered \emph{unlatched}, where the system can frequently switch between the output of the RTA filter and primary controller. In the event that \textit{Plan B} is insufficient, a fault compromises its safety assurance, or a human operator wants to ``pause" operations until they can investigate anomalous behavior, a \textit{Plan C} RTA filter may switch to the backup controller and remained \emph{latched} for a longer time period until a specified condition is met. A \textit{Plan D} RTA filter uses a similar \emph{latched} approach, where the goal is to immediately act to prevent complete loss. \textit{Plan B} RTA filters are the least invasive to the primary controller while still assuring safety, while \textit{Plan C} and \textit{Plan D} RTA filters provide increasing degradation in the case of anomalous system behavior. This paper focuses on developing \textit{Plan B} and \textit{Plan C} RTA approaches as \textit{Plan D} approaches, such as sun safe mode \cite{hoffman1999near,burk2010managing}, already exist. 

Although not always referred to as RTA, the concept of online safety assurance has been used in several applications such as autonomous vehicles \cite{hu2019lane}, 
fixed wing aircraft collision avoidance \cite{squires2019composition}, 
spacecraft collision avoidance \cite{mote2021natural}, VTOL aircraft \cite{singletary2020safety}, 
and many more. RTA has also been compared during Reinforcement Learning Training for 2D spacecraft docking \cite{dunlap2022run}.


\section{RUN TIME ASSURANCE}

Safety-critical systems may be modeled as dynamical systems, where $\boldsymbol{x} \in \mathbb{R}^n$ denotes the state vector and $\boldsymbol{u}\in \mathbb{R}^m$ denotes the control vector. Assuming control affine dynamics, a continuous-time system model is given by a system of ordinary differential equations where,
\begin{equation} \label{eq:fxgu}
   \boldsymbol{\dot{x}} = f(\boldsymbol{x}) + g(\boldsymbol{x})\boldsymbol{u}.
\end{equation}

\begin{figure}[tb!]
    \centering
    \includegraphics[width=0.9\columnwidth]{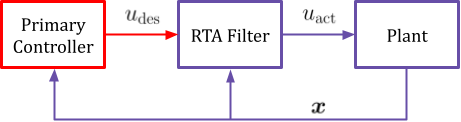}
    \caption{Feedback Control System with RTA.}
    \label{fig:RTA}
\end{figure}

At their core, RTA systems decouple the task of assuring safety from all other objectives of the controller. The control system is split into a performance-driven primary controller and a safety-driven RTA filter, which allows the designer to isolate unverified components of the system. This concept is shown in Figure \ref{fig:RTA}, where components with low safety confidence are outlined in red and components with high safety confidence are outlined in blue. In a typical feedback control system, the controller directly interacts with the plant. In a feedback control system with RTA, the RTA filter preempts potentially unsafe inputs from the primary controller, referred to as the desired input $u_{\rm des}$, and instead outputs a safe action, referred to as the actual input $u_{\rm act}$. The RTA filter intervenes when necessary to enforce safety, and otherwise allows $u_{\rm des}$ to pass through unaltered.

\subsection{Defining Safety}

To assure safety of a dynamical system, inequality constraints $\varphi_i(\boldsymbol{x}): \mathbb{R}^n\to \mathbb{R}$,  $\forall i \in \{1,...,M\}$ are defined for $M$ safety constraints on the state vector, such that $\varphi_i(\boldsymbol{x}) \geq 0$ when the constraint is satisfied. The allowable set $\mathcal{C}_{\rm A}$, defined as the set of states that satisfies all of the safety constraints, is given by,
\begin{equation}
    \mathcal{C}_{\rm A} := \{\boldsymbol{x} \in \mathbb{R}^n | \varphi_i(\boldsymbol{x}) \geq 0, \forall i \in \{1,...,M\} \}.
\end{equation}

In real world systems, constraints on actuation further limit safety to a forward invariant subset $\mathcal{C}_{\rm S} \subseteq \mathcal{C}_{\rm A}$ referred to as the safe set, where
\begin{equation}
    \boldsymbol{x}(t_0) \in \mathcal{C}_{\rm S} \Longrightarrow \boldsymbol{x}(t) \in \mathcal{C}_{\rm S}, \forall t \geq t_0.
\end{equation}

This set is said to be control invariant if a control law $\boldsymbol{u}$ exists that is forward invariant given constraints on actuation. 
Because the methods used to obtain $\mathcal{C}_{\rm S}$ are generally not scalable to complex and higher order systems, 
it is helpful to define a forward invariant backup set $\mathcal{C}_{\rm B} \subseteq \mathcal{C}_{\rm S}$ that can be more easily defined based on the design requirements.

\subsection{Switching-based Algorithms}

Switching-based RTA filters contain a monitor to the desired input $u_{\rm des}$ from the primary controller for safety. If $u_{\rm des}$ is safe, it is passed unaltered to the plant as $u_{\rm act}$; otherwise, a backup input $u_{\rm b}$ from a verified backup controller will be substituted for $u_{\rm act}$ passed to the plant. One possible implementation of a switching-based RTA filter is constructed as follows.

\noindent \rule{1\columnwidth}{0.7pt}
\noindent \textbf{Switching Filter}
\begin{equation}
\begin{array}{rl}
u_{\text{act}}(\boldsymbol{x})=
\begin{cases} 
u_{\rm des}(\boldsymbol{x}) & {\rm if}\quad \phi_1^{u_{\rm des}}(\boldsymbol{x}) \in \mathcal{C}_{\rm S}  \\ 
u_{\rm b}(\boldsymbol{x})  & {\rm if}\quad otherwise
\end{cases}
\end{array}\label{eq:switching}
\end{equation}
\noindent \rule[7pt]{1\columnwidth}{0.7pt}

Here, $\phi_1^{u_{\rm des}}(\boldsymbol{x})$ represents a prediction of the state $\boldsymbol{x}$ if $u_{\rm des}$ is applied for one discrete time interval.

The difference between explicit and implicit switching filters is the way in which $\mathcal{C}_{\rm S}$ is defined. $\mathcal{C}_{\rm S}$ can be defined \textit{explicitly},
\begin{equation} \label{eq: explicitly_defined_safe_set}
    \mathcal{C}_{\rm S} = \{\boldsymbol{x}\in\mathbb{R}^n \, | \, h(\boldsymbol{x})\geq 0 \},
\end{equation} 
with inequality constraint function $ h(\boldsymbol{x}):\mathbb{R}^n\to\mathbb{R}$, such that verifying $\boldsymbol{x}\in \mathcal{C}_{\rm S}$ is equivalent to verifying $h(\boldsymbol{x})\geq0$.
%
$\mathcal{C}_{\rm S}$ can also be defined \textit{implicitly} using closed loop trajectories under a backup control law,
\begin{equation} 
    \mathcal{C}_{\rm S} = \{\boldsymbol{x}\in\mathbb{R}^n \, | \, \forall t\geq 0,\,\, \phi^{u_{\rm b}}(t;\boldsymbol{x})\in\mathcal{C}_{\rm A} \} 
\end{equation}
where $\phi^{u_{\rm b}} $ represents a prediction of the state $\boldsymbol{x}$ for $t$ seconds under $u_{\rm b}$. 


\subsection{Optimization-based Algorithms}

Optimization-based RTA algorithms use gradient computations to create a set of barrier constraints, as presented in \cite{ames2019control, gurriet2020scalable, gurriet2018online, chen2021backup}. By optimizing the actual input $u_{\rm act}$, the algorithm is minimally invasive with respect to the safety constraints and results in smoother intervention when compared to a switching-based approach. Optimization filters can use a quadratic program, where the objective function is minimizing the $l^2$ norm difference between $u_{\rm des}$ and $u_{\rm act}$. One possible implementation of the optimization-based RTA filter is constructed as follows.

\noindent \rule{1\columnwidth}{0.7pt}
\noindent \textbf{Optimization Filter}
\begin{equation}
\begin{split}
u_{\text{act}}(\boldsymbol{x})={\text{argmin}} & \left\Vert \boldsymbol{u}_{\text{des}}-\boldsymbol{u}\right\Vert ^{2}\\
\text{s.t.} \quad & BC_i(\boldsymbol{x},\boldsymbol{u})\geq 0, \quad \forall i \in \{1,...,M\}
\end{split}\label{eq:optimization}
\end{equation}
\noindent \rule[7pt]{1\columnwidth}{0.7pt}

Here, $BC_i(\boldsymbol{x},\boldsymbol{u})$ represents a set of $M$ barrier constraints used to enforce safety of the system. The purpose of these constraints is to enforce Nagumo's condition \cite{nagumo1942lage}. 
To apply Nagumo's condition, the boundary of the set formed by $h(\boldsymbol{x})$ is examined to ensure $\dot{h}(\boldsymbol{x})$ is never decreasing. This condition can be written as,
\begin{equation}
    \dot{h}(\boldsymbol{x}) = \nabla h(\boldsymbol{x}) \dot{\boldsymbol{x}} = L_f h(\boldsymbol{x}) + L_g h (\boldsymbol{x}) \boldsymbol{u} \geq 0
\end{equation}
where $L_f$ and $L_g$ are Lie derivatives of $f$ and $g$ respectively. Since the boundary of the set has no volume, it is not practical to enforce this condition on its own. Instead, a class $\kappa$ strengthening function $\alpha$ is used to enforce the constraints on the boundary and relax the constraints away from the boundary. The barrier constraint is then written as,
\begin{equation}
    BC(\boldsymbol{x},\boldsymbol{u}) := L_f h(\boldsymbol{x}) + L_g h (\boldsymbol{x}) \boldsymbol{u} + \alpha(h(\boldsymbol{x})) \geq 0.
\end{equation}

Similarly to switching-based filters, the difference between explicit and implicit optimization filters is how these barrier constraints are defined. Using the system model given in (\ref{eq:fxgu}), the barrier constraints can be defined \textit{explicitly} as,
\begin{multline} \label{eq:exp_BC}
    BC_i(\boldsymbol{x},\boldsymbol{u}) := \nabla h_i(\boldsymbol{x}) (f(\boldsymbol{x}) + g(\boldsymbol{x})\boldsymbol{u}) + \alpha(h_i(\boldsymbol{x})) \geq 0, \\  \quad \forall i \in \{1,...,M\}.
\end{multline}
Again, $h_{i}(\boldsymbol{x})$ is a set of $M$ control invariant safety constraints. Explicit optimization methods differ from switching-based methods in that they do not need a backup controller. 
The barrier constraints can also be defined \textit{implicitly} as,
\begin{multline} \label{eq:imp_BC}
    BC_i(\boldsymbol{x},\boldsymbol{u}) := \nabla \varphi_i(\phi^{u_b}_j) D(\phi^{u_b}_j) [f(\boldsymbol{x}) + g(\boldsymbol{x})\boldsymbol{u}-f(\phi^{u_b}_j) \\ -g(\phi^{u_b}_j)u_{\rm b}(\phi^{u_b}_j)] + \alpha(\varphi_i(\phi^{u_b}_j) ), \quad \forall i \in \{1,...,M\}
\end{multline}
where $\varphi_i(\boldsymbol{x})$ is a set of $M$ safety constraints that define the allowable set $\mathcal{C}_{\rm A}$, $\phi^{u_b}_j$ refers to the $j^{th}$ discrete time interval along the backup trajectory $\forall t \in [0,T)$, and $D(\phi^{u_b}_j)$ is computed by integrating a sensitivity matrix along the backup trajectory. Note that for practical implementation, the trajectory is evaluated over a finite set of points. Implicit optimization filters typically use the same verified backup controller as implicit switching filters, which would be suitable for degradation to a \textit{Plan C} RTA system.
Implicit methods also introduce a trade-off between computation time and safety guarantees, as when more samples are taken along the trajectory, the computation time increases while safety can be guaranteed over a longer time horizon.

\section{SPACECRAFT DOCKING PROBLEM}

In the spacecraft docking problem, an active ``deputy" spacecraft approaches a passive ``chief" spacecraft to simulate docking in a linearized relative motion reference frame. Both spacecraft are assumed to be rigid bodies, represented as point-mass objects. It is also assumed that the mass of both satellites is significantly smaller than the mass of Earth, mass loss during maneuvers is significantly smaller than the spacecraft mass, the chief spacecraft is in a circular orbit, and that the distance between the spacecraft is significantly smaller than the distance of either spacecraft to Earth. This section discusses the dynamics, safety constraints, and applications of RTA for spacecraft docking.

\subsection{Dynamics}

\begin{figure}[hbt!]
\centering
\includegraphics[width=.35\textwidth]{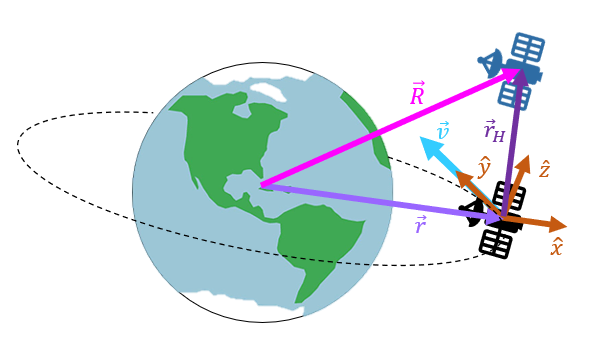}
\caption{Hill's reference frame centered on a chief spacecraft.}
\label{fig:HillsFrame}
\end{figure}

The location of the deputy with respect to the chief is expressed in Hill's reference frame \cite{hill1878researches}, where the origin is located at the mass center of the chief. As shown in Figure \ref{fig:HillsFrame}, the vector $\hat{x}$ points away from the center of the Earth, the vector $\hat{y}$ points in the direction of motion, and the vector $\hat{z}$ is normal to $\hat{x}$ and $\hat{y}$.

A first order approximation of the relative motion dynamics between the deputy and chief spacecraft is given by the Clohessy-Wiltshire equations \cite{clohessy1960terminal}, 
\begin{equation} \label{eq: system dynamics}
    \dot{\boldsymbol{x}} = A {\boldsymbol{x}} + B\boldsymbol{u}
\end{equation}
where $\boldsymbol{x}=[x,y,z,\dot{x},\dot{y},\dot{z}]^T \in \mathcal{X}=\mathbb{R}^6$ is the state vector, $\boldsymbol{u}= [F_x,F_y,F_z]^T \in U = [-u_{\rm max},u_{\rm max}]^3$ is the control vector, and
\begin{align}
\centering
    A = 
\begin{bmatrix} 
0 & 0 & 0 & 1 & 0 & 0 \\
0 & 0 & 0 & 0 & 1 & 0 \\
0 & 0 & 0 & 0 & 0 & 1 \\
3n^2 & 0 & 0 & 0 & 2n & 0 \\
0 & 0 & 0 & -2n & 0 & 0 \\
0 & 0 & -n^2 & 0 & 0 & 0 \\
\end{bmatrix}, 
    B = 
\begin{bmatrix} 
 0 & 0 & 0 \\
 0 & 0 & 0 \\
 0 & 0 & 0 \\
\frac{1}{m} & 0 & 0 \\
0 & \frac{1}{m} & 0 \\
0 & 0 & \frac{1}{m} \\
\end{bmatrix}.
\end{align}

In these equations, $n=\sqrt{{\mu}/{a^3}}$ is the spacecraft mean motion and $m$ is the mass of the deputy.

\subsection{Defining the Allowable Set} \label{sec:allowable}

The system is defined to be safe if it follows two safety constraints for all time: a distance dependent speed limit and a maximum velocity limit. The distance dependent speed limit is defined as,
\begin{equation} \label{eq:Constraint 1}
    \Vert \boldsymbol{v}_{\rm H} \Vert \leq \nu_0 + \nu_1 \Vert \boldsymbol{r}_{\rm H} \Vert 
\end{equation}
where $\nu_0$ defines the maximum allowable docking velocity, $\nu_1$ is a constant, and
\begin{equation}
    \Vert {\boldsymbol{r}_{\rm H}}\Vert =(x^2+y^2+z^2)^{1/2}, \quad \Vert {\boldsymbol{v}_{\rm H}} \Vert =(\dot{x}^2+\dot{y}^2+\dot{z}^2)^{1/2}.
\end{equation}
This constraint forces the deputy to slow down as it approaches the chief to avoid crashing. As an inequality constraint, it can be written as,
\begin{equation} \label{eq:ha1}
    \varphi_1(\boldsymbol{x}) := \nu_0+\nu_1 \Vert {\boldsymbol{r}_{\rm H}}\Vert - \Vert {\boldsymbol{v}_{\rm H}} \Vert.
\end{equation}
The maximum velocity limit is defined as,
\begin{equation} \label{eq:Constraint 2}
    | \dot{x} | \leq v_{\rm max}, \quad | \dot{y} | \leq v_{\rm max}, \quad | \dot{z} | \leq v_{\rm max}
\end{equation}
where $v_{\rm max}$ is a constant given by,
\begin{equation}
    v_{\rm max} = \frac{u_{\rm max}}{m} (t_{\rm stop})
\end{equation}
and $t_{\rm stop}$ is the time required to reach zero velocity in the $\hat{x}$, $\hat{y}$, or $\hat{z}$ direction. As inequality constraints, the maximum velocity limit can be written as,

\begin{equation} \label{eq:ha2}
\begin{gathered}
    \varphi_2(\boldsymbol{x}) := v_{\rm max}^2 - \dot{x}^2, \quad \varphi_3(\boldsymbol{x}) := v_{\rm max}^2 - \dot{y}^2, \\
    \varphi_4(\boldsymbol{x}) := v_{\rm max}^2 - \dot{z}^2.
\end{gathered}
\end{equation}

The allowable set $\mathcal{C}_{\rm A}$ is then defined as the set of states where $\varphi_i \geq 0$ for all four constraints.

\subsection{Defining the Safe Set} \label{sec:safe}

Next, the safe set $\mathcal{C}_{\rm S}$ is determined given the constraint on actuation, $\boldsymbol{u} \in U = [-u_{\rm max},u_{\rm max}]^3$.

\begin{assump} \label{assump:Rmax}
    $\Vert \boldsymbol{r}_{\rm H} \Vert$ is upper bounded by a maximum distance $R_{\rm max}$.
\end{assump}

This assumption is due to the fact that when $\Vert \boldsymbol{r}_{\rm H} \Vert $ becomes large, the linearization used in obtaining the equations of motion given in (\ref{eq: system dynamics}) breaks down and is no longer valid.

\begin{lem} \label{lem:speedlim}
    A maximum control input of
    \begin{equation}
        u_{\rm max} \geq (3n^2+2n\nu_1 + \nu_{1}^{2})R_{max} + (2n+\nu_{1})\nu_{0}
    \end{equation}
    guarantees that $h_{1}(\boldsymbol{x}) = \varphi_1(\boldsymbol{x})$.
\end{lem}

\noindent \textbf{Proof:} Assume that the system is at a point on the boundary of $\varphi_1(\boldsymbol{x})$, i.e., $\varphi_1(\boldsymbol{x})=0$. Since the environment is spherically symmetric, the problem can be considered in polar coordinates, where only the radial component of the velocity affects whether the constraint is satisfied at a future instant in time. Denote the radial and tangential components of the velocity vector as $v_{r}$ and $v_{\theta}$, respectively. Note that $v_{\theta}$ does not affect the radial component of the position, $r_r$, and only causes the point to travel along the level set $\varphi_1(\boldsymbol{x})= 0$. 

A lower bound on $u_{\rm max}$ is defined that guarantees the control invariance of $\mathcal{C}_{\rm S}$ by inspecting the worst-case scenario. This scenario occurs when $\Vert \boldsymbol{r}_{\rm H} \Vert=R_{\rm max}$, $\Vert \boldsymbol{v}_{\rm H} \Vert=v_{\rm max}$, and $r_r$ and $-v_r$ are pointed in the $\hat{x}$ direction. Given the dynamics in (\ref{eq: system dynamics}), the worst-case acceleration occurs in the $\dot{x}$-component when $\boldsymbol{x}=[-R_{\rm max},0,0,0,-v_{\rm max},0]^T$, and thus,
\begin{equation}
    u_{r} = u_{\rm max} - 3n^{2}R_{\rm max} - 2nv_{\rm max}.
\end{equation}
In this case,
%
%
\begin{equation}
    \frac{d}{dt}(\Vert \boldsymbol{r}_{\rm H} \Vert) = -v_r, \quad \frac{d}{dt}(\Vert \boldsymbol{v}_{\rm H} \Vert) = u_r.
\end{equation}

To ensure safety, the following condition must hold (noting that both terms are negative),
\begin{align}
    \frac{d}{dt}(\nu_{1}\Vert \boldsymbol{r}_{\rm H} \Vert) \leq \frac{d}{dt}(\Vert \boldsymbol{v}_{\rm H} \Vert) \implies \nu_{1}v_{r} \leq u_{r}.
\end{align}

Given Assumption \ref{assump:Rmax}, $v_{\rm max} = \nu_{1}R_{\rm max} + \nu_{0}$. By stating that $v_r=v_{\rm max}$,

\begin{align}
    \nu_{1}^{2}R_{max} + \nu_{1}\nu_{0} \leq u_{r}
\end{align}
where $u_r$ is the worst-case acceleration. 
By substituting and rearranging terms, it can be found that
$h_{1}(\boldsymbol{x})$ = $\varphi_1(\boldsymbol{x})$ if
\begin{equation}
    u_{\rm max} \geq (3n^2+2n\nu_1 + \nu_{1}^{2})R_{max} + (2n+\nu_{1})\nu_{0}. \; \Box
\end{equation}

\begin{lem} \label{lem:maxvel}
A maximum control input of
\begin{equation}
\begin{gathered}
    (3n^2x + 2n\dot{y})^2 < \left(\frac{u_{\rm max}}{m}\right)^2, \quad (-2n\dot{x})^2 < \left(\frac{u_{\rm max}}{m}\right)^2, \\ \quad (-n^2z)^2 < \left(\frac{u_{\rm max}}{m}\right)^2,
\end{gathered}
\end{equation}
guarantees that $h_{2}(\boldsymbol{x}) = \varphi_2(\boldsymbol{x})$, $h_{3}(\boldsymbol{x}) = \varphi_3(\boldsymbol{x})$, and $h_{4}(\boldsymbol{x}) = \varphi_4(\boldsymbol{x})$ respectively.
\end{lem}

\noindent \textbf{Proof:} From (\ref{eq: system dynamics}), in order for the safety constraints to be control invariant, the magnitude of acceleration from the term $B\boldsymbol{u}$ must be greater than the magnitude of acceleration from the term $A\boldsymbol{x}$ to ensure the velocity components of the state vector $\boldsymbol{x}$ are controllable. This becomes,

\begin{equation}
\begin{gathered}
    (3n^2x + 2n\dot{y})^2 < \left(\frac{u_{\rm max}}{m}\right)^2, \quad (-2n\dot{x})^2 < \left(\frac{u_{\rm max}}{m}\right)^2, \\ \quad (-n^2z)^2 < \left(\frac{u_{\rm max}}{m}\right)^2
\end{gathered}
\end{equation}
for $\dot{x}$, $\dot{y}$, and $\dot{z}$ respectively. $\Box$

\begin{assump} \label{assump:values}
    For this paper, $u \in [-1,1]^3$ N, $n = 0.001027$ rad/s, $m = 12$ kg, $\nu_0 = 0.2$ m/s, $\nu_1=4n$, $v_{\rm max} = 10$ m/s, and $R_{\rm max} = 10$ km.
\end{assump}

$\mathcal{C}_{\rm S}$ is dependent on a specific control law and system dynamics, and therefore it varies depending on the system. Given Assumption \ref{assump:values}, both Lemma \ref{lem:speedlim} and Lemma \ref{lem:maxvel} are valid, and therefore $\mathcal{C}_{\rm S}=\mathcal{C}_{\rm A}$ and $\mathcal{C}_{\rm S}$ is control invariant.

\subsection{Defining the Backup Set}

Solutions to (\ref{eq: system dynamics}) where $\boldsymbol{u}=0$ are known as natural motion trajectories (NMTs) \cite{frey2017constrained}. Elliptical closed NMTs centered at the origin are periodic solutions that also satisfy,

\begin{equation}
    \Dot{y}(0)=-2nx(0), \quad \dot{x}(0)= \frac{n}{2}y(0).
\end{equation}

\begin{figure}[htb!]
    \centering
    \includegraphics[width=0.5\columnwidth]{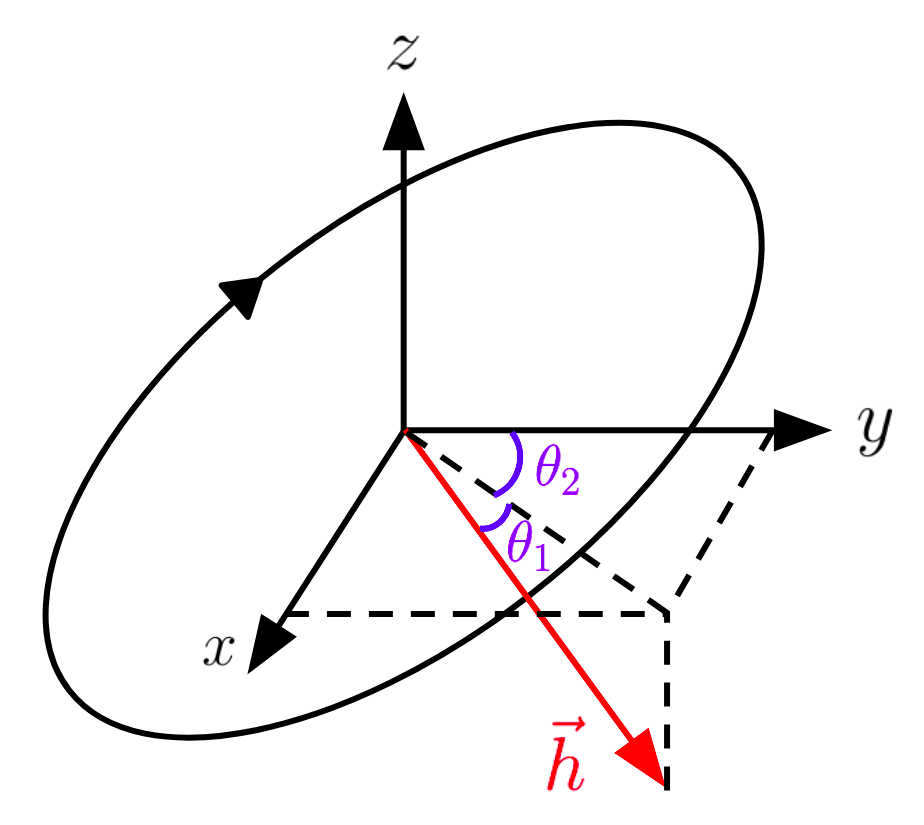}
    \caption{Depiction of angles $\theta_1$ and $\theta_2$ for an elliptical NMT.}
    \label{fig:theta}
\end{figure}

Elliptical NMTs are useful because they provide convenient ``parking orbits" in the event of a fault, where the spacecraft will stay on the NMT for all time without using fuel. Note that since elliptical NMTs are a set of closed trajectories, they create an invariant set. Additionally, all elliptical NMTs satisfy the following equations,
\begin{equation}
\label{eq:ellipse}
\begin{gathered}
    x(0)=b \sin (\nu), \quad \Dot{x}(0)=bn \cos (\nu), \\
    y(0)=2b \cos (\nu), \quad \Dot{y}(0)=-2bn \sin (\nu), \\
    z(0)=c \sin (\psi), \quad \Dot{z}(0)=nc \cos (\psi), \\
\end{gathered}
\end{equation}
where,
\begin{equation}
\begin{gathered}
\label{Eq:c}
    c = \frac{b}{\sin \theta_1} \sqrt{\tan^2 \theta_2 + 4 \cos^2 \theta_1}, \\
    \nu = \tan^{-1} \left(2 \frac{\cos \theta_1}{\tan \theta_2}\right)-\psi.
\end{gathered}
\end{equation}
In these equations, $b$ is the semi-minor axis, $\theta_1$ and $\theta_2$ are angles from the $x-y$ plane and the $y-z$ plane to the angular momentum vector as shown in Figure \ref{fig:theta}, and $\psi$ is the phase angle. To ensure that $\mathcal{C}_{\rm B}\subseteq\mathcal{C}_{\rm S}$, it must be proven that $\mathcal{C}_{\rm S}$ contains $\mathcal{C}_{\rm B}$.

\begin{lem}
$\mathcal{C}_{\rm B}\subseteq\mathcal{C}_{\rm S}$ if $\mathcal{C}_{\rm B}$ contains a set of elliptical NMTs adhering to the constraints,

\begin{equation}
\begin{gathered}
    \frac{\tan^2 \theta_2 + 4 \cos^2 \theta_1}{\sin^2 \theta_1} \leq \left(\frac{\nu_1}{n}\right)^2-4, \\
    b \leq \frac{v_{\rm max}}{2n}, \quad c \leq \frac{v_{\rm max}}{n},
\end{gathered}
\end{equation}
\end{lem}

\noindent \textbf{Proof:} Consider the point on an elliptical NMT where $\Vert \boldsymbol{r}_{\rm H} \Vert$ is minimized and $\Vert \boldsymbol{v}_{\rm H} \Vert$ is maximized. If $h_{1}(\boldsymbol{x})\geq 0$ at this point, this condition will hold for the entire NMT. At this point, $\nu = \frac{\pi}{2}$ and $\psi = 0$. Evaluating (\ref{eq:ellipse}) at this point and assuming $\nu_0=0$, the constraint becomes,

\begin{equation}
    \sqrt{(2bn)^2+(nc)^2} \leq \nu_1 \sqrt{(b)^2}
\end{equation}
which can be simplified to,

\begin{equation}
    c^2 \leq b^2 \left[\left(\frac{\nu_1}{n}\right)^2-4\right].
\end{equation}

Substituting $c$ and simplifying:

\begin{equation}
    \frac{\tan^2 \theta_2 + 4 \cos^2 \theta_1}{\sin^2 \theta_1} \leq \left(\frac{\nu_1}{n}\right)^2-4.
\end{equation}

Next, consider the point along each axis where the elliptical NMT reaches its maximum velocity. If (\ref{eq:Constraint 2}) is not violated at these points, then it will never be violated for the entire NMT. These points are, $\dot{x} = bn$, $\dot{y} = 2bn$, and $\dot{z} = cn$. The constraint then becomes,

\begin{equation} \label{eq:Constraint 3}
    bn \leq v_{\rm max}, \quad 2bn \leq v_{\rm max}, \quad cn \leq v_{\rm max}
\end{equation}
and therefore,

\begin{equation}
    b \leq \frac{v_{\rm max}}{2n}, \quad c \leq \frac{v_{\rm max}}{n}. \; \Box
\end{equation}

\subsection{Control Law to Safe Backup Set} \label{sec:ub}

An LQR tracking controller is used to guide the spacecraft to the backup set, similar to the method used in \cite{frey2017constrained}. The controller first computes a discrete set of NMTs that are part of $\mathcal{C}_{\rm B}$, then guides the spacecraft to the closest point based on position to one of the NMTs. Once the $l^2$ norm between the spacecraft position and desired position is less than a set value $\epsilon$, the desired position begins to update based on the dynamics of the system. Since the desired point starts on an NMT, it will stay on the NMT for all time.

Both the implicit switching and implicit optimization RTA filters use this LQR controller as a backup controller to compute the backup trajectory. The explicit switching filter uses a simpler backup controller, where $u_{\rm b}$ is determined to be the value for $\boldsymbol{u}$ that causes $h_{i}(\boldsymbol{x})=0$ when the $i^{th}$ constraint is violated. Since multiple constraints can be violated at once, the RTA must have a hierarchy to determine which constraints take precedence. In this case, the distance dependent speed limit is given higher priority and is therefore evaluated second within $u_{\rm b}$.


The four RTA filters developed in this paper are all \textit{Plan B} RTA approaches, where the system can frequently switch between the RTA filter output and the \textit{Plan A} primary controller to assure safety in real time. The implicit methods use a \textit{Plan C} backup controller, which guides the spacecraft to a known invariant safe state along an elliptical NMT. Although not used for this problem, an example of a \textit{Plan D} RTA filter for spacecraft docking is ``sun safe mode," where the spacecraft points its antenna at the Earth and solar panels at the Sun to prevent complete loss \cite{hoffman1999near}.

\section{SIMULATION RESULTS}


An unconstrained LQR controller is used to guide the deputy spacecraft to the origin to simulate docking. This controller is aggressive and does not consider safety constraints, and is therefore a good example to prove the effectiveness of the four RTA approaches.
Each simulation is initialized with the same initial conditions to directly compare the results, where $\Vert \boldsymbol{r}_{\rm H} \Vert = 9.85$ km, $\Vert \boldsymbol{v}_{\rm H} \Vert = 0.866$ m/s, and the position of the chief is $[x,y] = [0,0]$. Euler integration with a time step of 1 second is used to simulate the dynamic system. For the implicit methods, the backup trajectories are evaluated over a period of 5 seconds.

\subsection{Results}

\begin{figure}[tb!]
    \begin{subfigure}[t]{0.49\columnwidth}
        \centering
        \includegraphics[width=\linewidth]{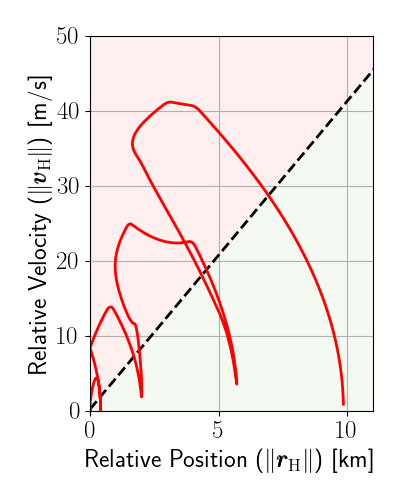}
        \caption{Dist. Dependent Speed Limit.}
        \label{fig:noRTA1}
    \end{subfigure}
    \begin{subfigure}[t]{0.49\columnwidth}
        \centering
        \includegraphics[width=\linewidth]{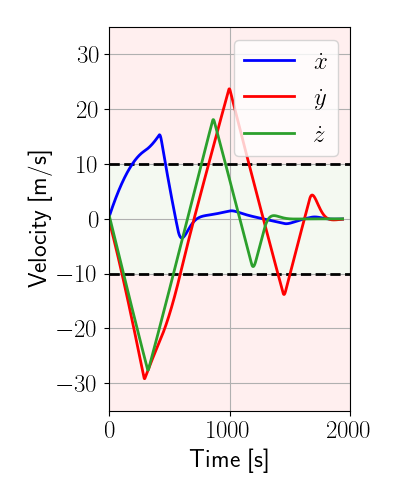}
        \caption{Maximum Velocity Limit.}
        \label{fig:noRTA2}
    \end{subfigure}
    \caption{Simulation Results with No RTA.}
    \label{fig:noRTA}
\end{figure}

This section shows the results of one simulation for each of the four RTA filters. In these figures, the green shaded region represents $\mathcal{C}_{\rm S}$, the red shaded region represents all unsafe states, and the black dotted lines represent where $h_{i}(\boldsymbol{x})=0$. Figure \ref{fig:noRTA} shows the results of a simulation where no RTA is used to demonstrate the unsafe nature of the primary controller.

\begin{figure}[tb!]
    \begin{subfigure}[t]{0.49\columnwidth}
        \centering
        \includegraphics[width=\linewidth]{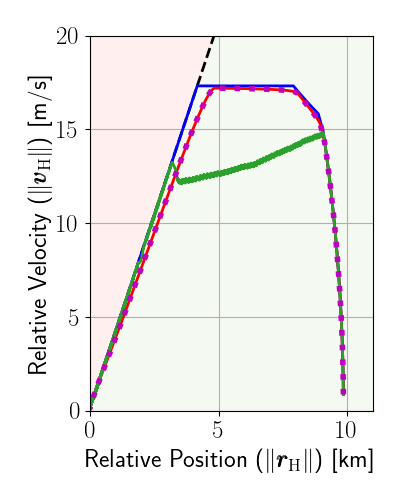}
        \caption{Dist. Dependent Speed Limit.}
        \label{fig:ImO1}
    \end{subfigure}
    \begin{subfigure}[t]{0.49\columnwidth}
        \centering
        \includegraphics[width=\linewidth]{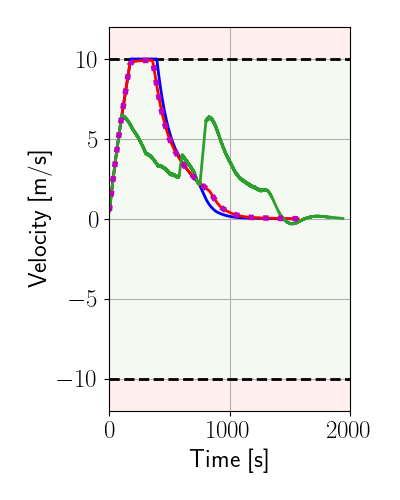}
        \caption{Maximum $\dot{x}$ Limit.}
        \label{fig:ImO2}
    \end{subfigure}
    \begin{subfigure}[t]{0.49\columnwidth}
        \centering
        \includegraphics[width=\linewidth]{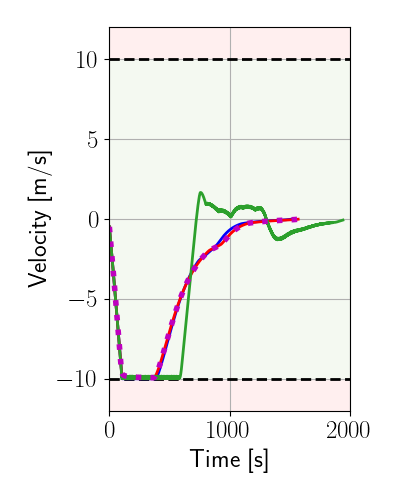}
        \caption{Maximum $\dot{y}$ Limit.}
        \label{fig:ImO3}
    \end{subfigure}
    \begin{subfigure}[t]{0.49\columnwidth}
        \centering
        \includegraphics[width=\linewidth]{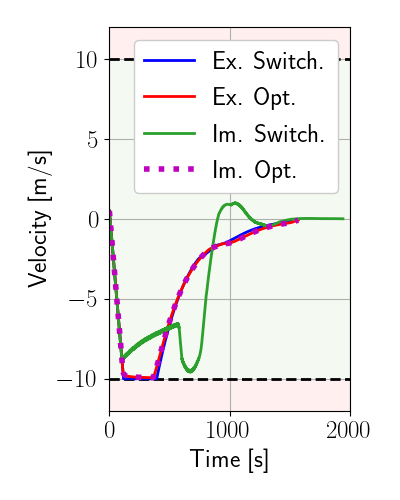}
        \caption{Maximum $\dot{z}$ Limit.}
        \label{fig:ImO4}
    \end{subfigure}
    \caption{Comparison of RTA Approaches.}
    \label{fig:sim}
\end{figure}

Figure \ref{fig:sim} shows the simulation results for all four RTA filters. For the explicit switching filter, when the primary controller is pushing the system into an unsafe state, RTA intervenes to ensure $h_{i}(\boldsymbol{x})=0$. For the explicit optimization filter, RTA begins to intervene as the system approaches the boundary of $\mathcal{C}_{\rm S}$, and prevents it from ever touching the  boundary. For the implicit switching filter, there is undesirable ``chattering," where the filter is frequently switching between the primary and backup controllers. Lastly, the behavior of the implicit optimization filter is shown to be visually identical to the explicit optimization filter.

Finally, the computation time for each approach is compared in Table \ref{tab:computation}. The average computation time over 100 simulations is calculated and then normalized by the smallest time (explicit switching) for comparison. 

\begin{table}[h]
\caption{Comparison of Computation Time.}
\label{tab:computation}
\begin{center}
\begin{tabular}{|c|c|c|}
\hline
RTA Filter & Time Multiple & Standard Deviation \\ \hline
Explicit Switching & 1.00x & 0.03 \\
Explicit Optimization & 3.25x & 0.13 \\
Implicit Switching & 11.31x & 1.05 \\
Implicit Optimization & 13.33x & 0.52 \\ \hline
\end{tabular}
\end{center}
\end{table}

\section{CONCLUSION}
This paper presented four approaches to run time assurance: explicit switching, implicit switching, explicit optimization, and implicit optimization. These approaches were compared through application on a safe satellite docking problem. 
Due to their faster computation times, explicit approaches are recommended if a control invariant set can be defined. However, explicit safe sets can be difficult to define as complexity of system dynamics and safety constraints increases, which make implicit approaches a popular choice. Switching approaches are recommended for use in processing-constrained and faster-than-real-time applications due to their faster computation time. Optimization approaches are recommended when computationally tractable for physical testing and operational use due to their minimal interference with the primary controller and their ability to better handle multiple constraints. Overall, all four approaches effectively assure safety in real time.





\bibliographystyle{Bibliography/IEEEtran}
\bibliography{Bibliography/IEEEabrv,Bibliography/root}

\end{document}